
\magnification=1200
\hsize 15.0 true cm
\vsize 23.0 true cm

\def\c{\centerline}
\def\v{\vskip 1pc}
\def\ej{\vfill\eject}
\def\r{\vec r}
\def\R{\vec R}
\def\P{\vec P}
\def\S{\vec S}
\def\L{\vec L}
\def\f{\phi}
\def\t{\vec \tau}
\def\ra{\rangle}
\def\la{\langle}
\def\e{{\rm e}}
\def\tr{{\rm tr}}
\def\half{{1 \over 2}}
\def\da{\dagger}
\def\eps{\epsilon_6}
\def\s{\vec \sigma}
\def\om{\omega}
\def\G{\Gamma}
\overfullrule 0 pc
\
\vskip 1.5pc
\parindent 3pc\parskip 1pc
\c{\bf An attractive nucleon-nucleon spin-orbit force from skyrmions with
dilatons}
\vskip 4 pc
\c{G. K\"albermann$^{\rm a}$ and J.M. Eisenberg$^{\rm b}$}
\v
\c{$^{\rm a}$ {\it Rothberg School for Overseas Students and Racah Institute
of Physics}}
\c{\it Hebrew University, 91904 Jerusalem, Israel}
\vskip 0.5 pc
\c{$^{\rm b}$ {\it School of Physics and Astronomy}}
\c{\it Raymond and Beverly Sackler Faculty of Exact Sciences}
\c{\it Tel Aviv University, 69978 Tel Aviv, Israel}

\vskip 4pc
\noindent {\bf Abstract:-}  Within the skyrmion approach for the
nucleon-nucleon force, difficulties have been experienced in obtaining an
isoscalar attractive spin-orbit potential, in parallel to the problems
of finding attraction in the
isoscalar central potential.  We here study the spin-orbit force
using a skyrmion with four- and six-derivative stabilizing terms in the
lagrangian as well as with the crucial addition of a dilaton.  With these
features present the spin-orbit force proves to be attractive as does the
central potential.  In the absence of the dilaton,
attraction can also be found for the spin-orbit potential but only at
the expense of a greatly over-emphasized term with six derivatives and
a continuing absence of attraction in the central potential.

\vfill

\noindent January, 1995.

\ej

\baselineskip 15 pt \parskip 1 pc \parindent 3pc

In the mid-1980s there was a great revival of interest in Skyrme's original
idea [1] whereby baryon physics is approximated by topological solitons.
Very soon thereafter a number of attacks were made on the two-nucleon
problem using skyrmions.  The general skyrmion approach has been
reviewed extensively [2--13] and particular emphasis on studies of the
$NN$ system is given in refs. [3,4,8--13].  The key difficulty in
describing the $NN$ force using skyrmions proves to be the subtle issue
of finding attraction in the isoscalar central potential.  This does not
easily emerge [3,4,8--12] from the
simplest approaches based on the product ansatz in which it is assumed that
the solution for baryon number $B = 2$ is well approximated by the
product of two solutions for $B = 1.$  Eventually several possible
mechanisms yielding central attraction were found.  These included the
use of a full numerical solution for the $B = 2$ system [12,13] or
alternatively the admixture of higher resonances in the
projected state for each of the two baryons [9,10].  These two
approaches may not be that dissimilar since both attempt to include
significant distortion of one nucleon due to the presence of the other
as suggested also in calculations based on the nonrelativistic quark
model [14].  In the two cases, of course, quite different degrees of
freedom are chosen to express this situation.

A different mechanism through which attraction is obtained has to do
with the
coupling of a dilaton to the skyrmion.  It has long been known [15] that
an effective chiral lagrangian can be made to mimic the scale breaking of
quantum chromodynamics by introducing a dilaton field tailored to yield
the same trace anomaly as possessed by QCD.  It is plausible that the
introduction of such an additional scale will help towards the solution
of the problem of $NN$ central attraction since it serves to sharpen the
surface of the baryon described by the skyrmion.  This in turn prevents the
repulsive nature of the $B = 2$ system for small interbaryon separations
from overwhelming the possibility of attraction in the
range of 2 fm or so.  Thus there is appreciable central attraction for
skyrmions deriving from a lagrangian that is coupled to a dilaton
[16,17].

More recently considerable effort has been expended in studying the $NN$
spin-orbit force with skyrmions [18--23], in particular the isoscalar part
thereof.  Once again it has not proved possible to obtain the correct,
negative sign for this component with the simple product ansatz [19,20].
This led Riska and Schwesinger [21] to suggest that a term with
six derivatives often used to augment skyrmion stabilization [24--26]
may bring about an attractive isoscalar spin-orbit force, although an
unusually
large coupling constant for this term was required to accomplish this [23].
They also suggested that the inclusion of a dilaton field would increase
this attraction.  The present study is a detailed working out of that idea.

Our point of departure is the skyrmion lagrangian with four- and
six-derivative terms and dilaton coupling to produce scale invariance,
$$\eqalign{{\cal L} & = {\cal L}_{2\, {\rm dilaton}} +
 {\cal L}_{2\, {\rm skyrmion}} + {\cal L}_4 + {\cal L}_6 - V(\f) \cr
& = \e^{2\f}\bigg[\half\G_0^2\partial_\mu\f\partial^\mu\f
- {F_\pi^2 \over 16}\tr(L_\mu L^\mu)\bigg]
+{1 \over 32 e^2}\tr[L_\mu,\, L_\nu]^2 \cr
& - \eps^2 \e^{-2\f} B_\mu B^\mu
- {C_G \over 4}[1 + \e^{4\f}(4\f-1)].}\eqno(1)$$
Here $L_\mu \equiv U^\da\partial_\mu U$ and
$B^\mu \equiv -(\epsilon^{\mu\alpha\beta\gamma}/24\pi^2)
\tr(L_\alpha L_\beta L_\gamma),$
where $U(\r,t)$ is the unitary SU(2) chiral field, $F_\pi$ is the
pion decay constant (with experimental value 186 MeV), $e$ is
the Skyrme parameter, and $\eps$ is the coefficient of the
six-derivative repulsive term.  In ref. [25] identification is made between
${\cal L}_6$ and $\om NN$ coupling (through a term closely related to
${\cal L}_6$ but containing a single trace over six factors of  $L_\mu$
rather than the two traces over three  $L_\mu$s shown here) and it is
determined that $\eps = g_\om^2/2 m_\om^2,$ where the $\om NN$ coupling
constant is empirically [25] $g_\om^2/4\pi \sim 10,$ the value we have
taken here, and the
$\om$ mass is $m_\om = 782$ MeV.  The dilaton parameters [15] are chosen
as $\G_0 = 137$ MeV and $C_G = (121\ {\rm MeV})^4.$

The static $B = 1$ solution for eq. (1) is taken in the usual hedgehog
form $U_0(\r) = \exp[i \t \cdot \hat{\r} F(r)]$ and for the $B = 2$
system we use the product ansatz
$$U_{B=2}  = A_1(t)U_0(\r-\r_1(t))A_1^\da(t)\,
A_2(t)U_0(\r-\r_2(t))A_2^\da(t),\eqno(2)$$
where $\r_1(t) = \r + \R(t)/2$ and $\r_2(t) = \r - \R(t)/2$ so that
$\r_1(t) - r_2(t) \equiv \R(t)$ is the dynamical separation
between the two nucleons, and $A_1(t)$ and $A_2(t)$ are the collective
rotations of skyrmion 1 and 2 from which
projection onto nucleon states is generated.
The dilaton for $B = 2$ is taken to be additive as usual [16],
$$\f_{B=2}=\f_1+\f_2.\eqno(3)$$
Both the assumptions of a product ansatz for the skyrmion and of an
additive form for the dilaton are expected to break down for small
separation distances but this is not of overly great concern here since
we are interested in the range $R > 1$ fm for the central and
spin-orbit potentials.  The evaluation of $NN$
potentials proceeds as usual by calculating the energy of the $B = 2$
system from the insertion of eq. (2) into eq. (1),
projecting onto nucleon states, and identifying the relevant part of the
energy (e.g., in the cases shown here only isoscalar components are
considered).  For the central potential $V_{\rm C}$ it is of course
necessary to subtract the energy $V_{\rm C}(R \to \infty).$

Forms for the $NN$ isoscalar spin-orbit potentials from skyrmions
given in the literature have been contradictory, nor do we get complete
agreement with the reported results.  We therefore attempt to sketch
here a few key steps in the calculation.  We use the convenient form [19]
$$\dot U(\r - \r_k(t)) = A_k(t) \bigg(
 {i \over 2\lambda}\, [\t \cdot \vec j_k,\, U_0(\r - \r_k)] -
 {\vec p_k \over M} \cdot \nabla U_0(\r - \r_k) \bigg)A_k^\da(t),
\quad\quad k = 1,2,\eqno(4)$$
where $M$ and $\lambda$ are the skyrmion mass and moment of inertia
[2--13] trivially modified by the dilaton such that parts arising from
${\cal L}_{2\, {\rm skyrmion}}$ acquire a factor $\e^{2\f}$ and those
coming from ${\cal L}_6$ have $\e^{-2\f}.$  In eq. (4), $\vec j_k$ is the
angular momentum operator for nucleon $k$ (i.e., $\vec j_k \to \half\s_k$,
where $\s_k$ is the set of Pauli matrices for nucleon $k$).
We also are aided by the projection theorem [19]
$$\la N'|A_k \t A_k^\da|N \ra = -{1 \over 3} \la N'|\s_k
(\t \cdot \t_k)|N \ra, \quad\quad k = 1,2,\eqno(5)$$
in which the subscripted operators refer to the nucleon space while
$\t$ is the skyrmion SU(2) matrix.  Reference [19] also gives
explicitly the expansion of ${\cal L}_2$ and ${\cal L}_4$ in terms of
$\dot U_k$ and $\nabla U_k.$  For the {\it isoscalar} spin-orbit case the
overall traces over the combined space of the $B = 2$ skyrmion can be
broken down into separate traces for the $k = 1$ and $k = 2$ factors.
It is important to note that in the presence of the dilaton the final
result contains terms referring to a single nucleon of the form
$\s_1 \times \r_1 \cdot \vec p_1 \to {1 \over 2} \s_1 \cdot \L
(1 +2 z/R),$ after averaging over the azimuthal angle taken with respect
to the direction of $\R,$ and ``crossed'' terms
referring to both nucleons of the form $\s_1 \times \r_1 \cdot \vec
p_2 \to - {1 \over 2} \s_1 \cdot \L (1 +2 z/R),$ where $\L
\equiv \vec R \times \P$ with $\P \equiv (\vec p_1 - \vec p_2)/2,$
and $z$ is the component of the skyrmion
variable along the direction of $\vec R.$  One also encounters from
${\cal L}_4$ combinations like $\P \cdot (\r + \R/2) (\r + \R/2) \cdot
(\s_2 \times (\r - \R/2)) \to -{1 \over 2} (r^2 - z^2)
(\s_2 \cdot \L),$ where again the arrow indicates averaging over the
azimuthal angle.

With the above prescriptions the calculation of the potentials reduces to
straightforward, if somewhat tedious, calculations of traces.  Our
result for the isoscalar spin-orbit potential is then
$$V_{{\rm SO}, I = 0} = {\S \cdot \L \over \lambda M}
(V_2 + V_4 + V_6),\eqno(6)$$
where $\S \equiv {1 \over 2}(\s_1 + \s_2)$ and, in parallel to the
skyrmion components of the lagrangian in eq. (1),
$$V_2 = {F_\pi^2 \over 8} \int
\bigg(\e^{2(\f_1 + \f_2)} - \e^{2\f_1}\bigg) {s_1^2 \over r_1^2}
(1 + 2z/R) d\r,\eqno(7{\rm a})$$
with $s_k \equiv \sin F_k, \ k = 1, 2$;
$$V_4 = {1 \over 2 e^2} \int
\bigg[{s_1^2 \over r_1^2} \bigg({F'}_2^2 + {3 s_2^2 \over r_2^2}\bigg)
(1 + 2z/R)
+ {s_2^2 \over r_1^2 r_2^2} \bigg({F'}_1^2 - {s_1^2 \over r_1^2}\bigg)
(r^2 - z^2)\bigg]d\r,\eqno(7{\rm b})$$
and
$$V_6 = -\eps^2 \bigg[ \int \e^{-2(\f_1 + \f_2)} B_1^0 B_2^0 d\r -
\int \bigg(\e^{-2(\f_1 + \f_2)} - \e^{-2\f_1}\bigg)(B_1^0)^2
(1 + 2z/R),\eqno(7{\rm c})$$
where $B_k^0 \equiv -F'_k\sin^2 F_k/2\pi^2 r_k^2, \ k = 1, 2$
is the baryon density for nucleon 1 or 2.

The calculational procedure now consists in solving the equations of
motion derived from ${\cal L}$ of eq. (1) for hedgehog skyrmion and
dilaton profiles.  Parameters for the skyrmion are
restricted by requiring reasonable values for the nucleon and
$\Delta$ masses.  These profiles are then inserted into
eqs. (7) to produce the spin-orbit potential (or into the well-known
expressions for the central potential---see refs. [3,9]).  Results
are shown in figs. 1 through 5.  The first two of these show
attraction of the correct magnitude as given by potentials fitted
to the $NN$ scattering data both for the spin-orbit and the
central potentials.  Both
cases contain the six-derivative term ${\cal L}_6$ and dilaton coupling.
Without the dilaton present (figs. 3 and 4) no attraction is found for
$V_{\rm C},$ as is expected from previous work on the
$NN$ system with skyrmions [3,9].  Figure 4 does show $V_{\rm SO} < 0$
as found in the work of ref. [21].  As there, this is achieved by
exaggerating the importance of the six-derivative term in the lagrangian
${\cal L}_6,$ in our case by ignoring ${\cal L}_4$ altogether.  This
seems to us to be artificial in that one expects [27] from $\pi\pi$
scattering that $e \approx 5 \pm 2$; it also leaves unsolved the
sources of central attraction.  Figure 5 uses ${\cal L}_4,$ with
a smaller value of $e,$ and ${\cal L}_6$ and the dilaton; it fails to
show attraction for $V_{\rm SO}$.  In fact
for values of $e$ less than about 9 we do not achieve
$V_{\rm SO}(R) < 0$ in the range of $NN$ separations of relevance here,
$1 < R < 2$ fm.  Although $e \ge 9$ is somewhat outside the range
suggested by $\pi\pi$ scattering, the combination of the
${\cal L}_6$ term and dilaton coupling does yield a reasonable,
semiquantitative picture for the
main ingredients of the $NN$ force as derived from skyrmions.
\v\v
This research was supported in part by the Israel Science Foundation
and in part by the Yuval Ne'eman Chair in Theoretical Nuclear Physics
at Tel Aviv University.  It is a pleasure to acknowledge very useful
exchanges on the subject matter of this work with Abdellatif Abada.
\v\v\v\v

\c{\bf References}
\v
\baselineskip 12pt
\parskip 0pc
\parindent 1pc
\hangindent 2pc
\hangafter 10

\item{[1]}  T.H.R. Skyrme, Proc. Roy. Soc. London, Series A, 260 (1961)
127; 262 (1961) 237 and Nucl. Phys. 31 (1962) 556.
\v
\item{[2]}  A.P. Balachandran, Proc. 1985 Theoretical Advanced Study
Institute, M.J. Bowick and F. Gursey, eds. (World Scientific, Singapore,
1985).
\v
\item{[3]}  R. Vinh Mau, Acta Phys. Austriaca, Suppl. 27 (1985) 91.
\v
\item{[4]}  G. Holzwarth and B. Schwesinger, Rep. Prog. Phys. 49 (1986)
825.
\v
\item{[5]}  I. Zahed and G.E. Brown, Phys. Repts. 142 (1986) 1.
\v
\item{[6]}  U.-G. Meissner and I. Zahed, Adv. Nucl. Phys. 17 (1986) 143.
\v
\item{[7]}  U.-G. Meissner, Phys. Repts. 161 (1988) 213.
\v
\item{[8]}  E.M. Nyman and D.O. Riska, Int. J. Mod. Phys. A 3 (1988)
1535.
\v
\item{[9]}  J.M. Eisenberg and G. K\"albermann, Prog. Nucl. Part. Phys.
22 (1989) 1.
\v
\item{[10]}  J.M. Eisenberg and G. K\"albermann, Nucl. Phys. A 508 (1990)
395c.
\v
\item{[11]}  E.M. Nyman and D.O. Riska, Rept. Prog. Phys. 53 (1990)
1137.
\v
\item{[12]}  M. Oka and A. Hosaka, Annu. Rev. Nucl. Part. Sci. 42 (1992)
333.
\v
\item{[13]}  T.S. Walhout and J. Wambach, Int. J. Mod. Phys. E 1 (1992)
665.
\v
\item{[14]}  K. Maltman and N. Isgur, Phys. Rev. D 29 (1984) 952.
\v
\item{[15]}  J. Schechter, Phys. Rev. D 21 (1980) 3393;

H. Gomm, P. Jain, R. Johnson, and J. Schechter, Phys. Rev. D 33 (1986) 801.
\v
\item{[16]}  H. Yabu, B. Schwesinger, and G. Holzwarth, Phys. Lett. B
224 (1989) 25.
\v
\item{[17]}   K. Tsushima and D.O. Riska, Nucl. Phys. A560 (1993) 985.
\v
\item{[18]}  D.O. Riska and E.M. Nyman, Phys. Lett. B 183 (1987) 7.
\v
\item{[19]}  D.O. Riska and K. Dannbom, Phys. Scripta 37 (1988) 7.
\v
\item{[20]}  T. Otofuji, S. Saito, M. Yasuno, H. Kanada, and R. Seki,
Phys. Lett. B 205 (1988) 145.
\v
\item{[21]}  D.O. Riska and B. Schwesinger, Phys. Lett. B 229 (1989)
339.
\v
\item{[22]}  R.D. Amado, B. Shao, and N.R. Walet, Phys. Lett. B 314
(1993) 159; 324 (1994) 467.
\v
\item{[23]}  A. Abada, preprint, Laboratoire de Physique Nucl\'eaire,
Universit\'e de Nantes, LPN 94-01, February, 1994; hep-ph/9401341.
\v
\item{[24]}  G.S. Adkins and C.R. Nappi, Phys. Lett. B 137 (1984) 251.
\v
\item{[25]}  A. Jackson, A.D. Jackson, A.S. Goldhaber, G.E. Brown, and
L.C. Castillejo, Phys. Lett. 154B (1985) 101.
\v
\item{[26]}  G. K\"albermann, J.M. Eisenberg, and A. Sch\"afer, Phys.
Lett. B 339 (1994) 211;

G. K\"albermann and J.M. Eisenberg, preprint, Tel Aviv University,
January,

1995; hep-ph/9501302.
\v
\item{[27]}  J.F. Donoghue, E. Golowich, and B.R. Holstein, Phys. Rev.
Lett. 53 (1984) 747.
\v
\ej
\baselineskip 15 pt \parskip 1 pc \parindent 0 pc
 {\bf Figure captions:-}

Fig. 1.  Spin-orbit potential and central potential (dashed line for
$r \ge 2$ fm)
for skyrmion parameters $F_\pi = 143$ MeV and $e = 20.0,$ yielding a
nucleon mass $M_N = 942$ MeV and a $\Delta$ mass $M_\Delta = 1174$ MeV.
Here ${\cal L}_4,$ ${\cal L}_6,$ and the dilaton are all present and
both $V_{\rm SO}$ and $V_{\rm C}$ are attractive.

Fig. 2.  Spin-orbit and central potentials for $F_\pi = 140$ MeV and
$e = 9.0,$ yielding $M_N = 976$ MeV and $M_\Delta = 1169$ MeV.
Curves are as defined in fig. 1.

Fig. 3.  Spin-orbit and central potentials for $F_\pi = 112$ MeV and
$e = 4.84$ without dilaton coupling.  Here $M_N = 978$ MeV and
$M_\Delta = 1106$ MeV.  Curves are as defined in fig. 1.
Neither $V_{\rm SO}$ nor $V_{\rm C}$ is attractive.

Fig. 4.  Spin-orbit and central potentials for $F_\pi = 140$ MeV using
only ${\cal L}_6$ to stabilize the skyrmion
(i.e., ${\cal L}_4 \equiv 0$) and without dilaton coupling.  The mass
values that emerge are $M_N = 942$ MeV and $M_\Delta = 1251$ MeV.
Curves are as defined in fig. 1.

Fig. 5.  Spin-orbit and central potentials for $F_\pi = 112$ MeV and
$e = 4.84$ with ${\cal L}_4,$ ${\cal L}_6,$ and the dilaton all present,
yielding $M_N = 922$ MeV and $M_\Delta = 1048$ MeV.
Curves are as defined in fig. 1.

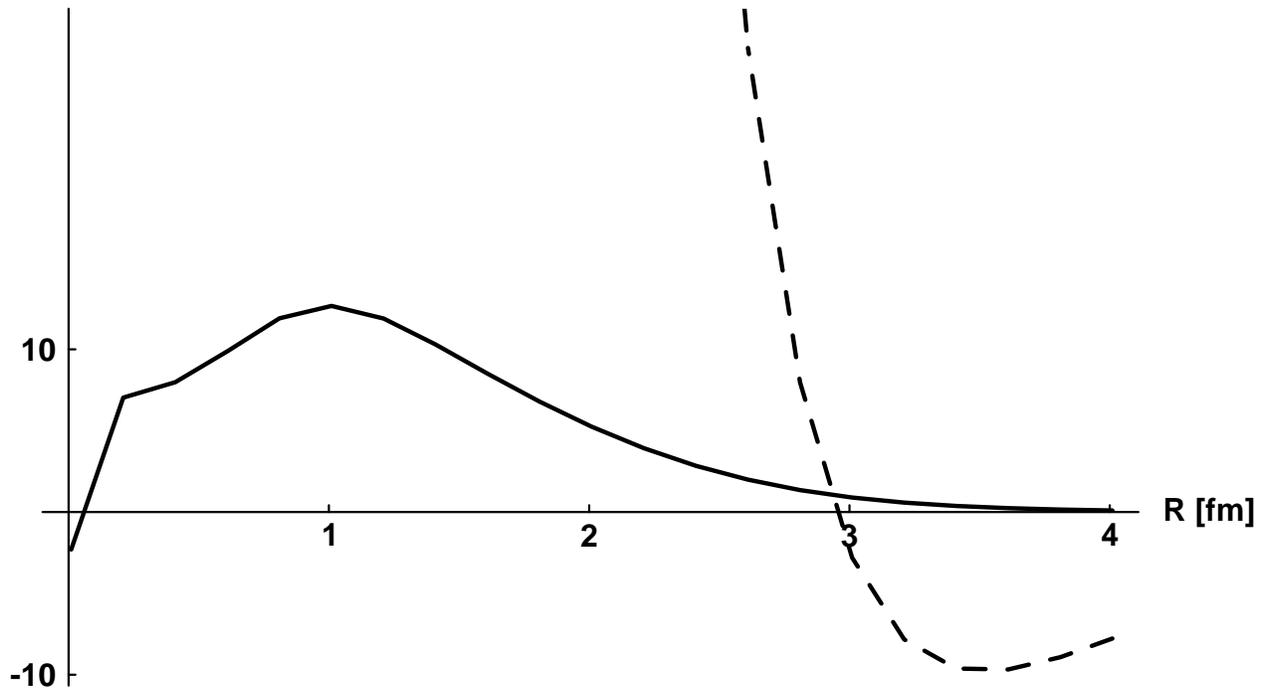
\bye